\documentclass[aps,prl,floatfix,showpacs,tightenlines,superscriptaddress]{revtex4}

\usepackage{amsmath}
\usepackage{graphicx}
\usepackage{epsfig}

 \newcommand{\ket}[1]{ {\left| #1 \right\rangle} }
 \newcommand*{\half}{\ensuremath{\frac{1}{2}}}

\begin{document}

\title{Applications of Electromagnetically Induced Transparency to Quantum Information Processing}

\author{R.\ G.\ Beausoleil}\email{ray.beausoleil@hp.com}
 \affiliation{Hewlett-Packard Laboratories,
13837 175$^\textrm{th}$ Pl.\ NE, Redmond, WA 98052--2180, USA}

\author{W.\ J.\ Munro}
\affiliation{Hewlett-Packard Laboratories, Filton Road, Stoke Gifford,
Bristol BS34 8QZ, United Kingdom}

\author{D.\ A.\ Rodrigues}
\affiliation{Hewlett-Packard Laboratories, Filton Road, Stoke
Gifford, Bristol BS34 8QZ, United Kingdom}

\author{T.\ P.\ Spiller}
\affiliation{Hewlett-Packard Laboratories, Filton Road, Stoke Gifford,
 Bristol BS34 8QZ, United Kingdom}

\date{February 15, 2004}

\begin{abstract}
We provide a broad outline of the requirements that should be met
by components produced for a Quantum Information Technology (QIT)
industry, and we identify electromagnetically induced transparency
(EIT) as potentially key enabling science toward the goal of
providing widely available few-qubit quantum information
processing within the next decade. As a concrete example, we build
on earlier work and discuss the implementation of a two-photon
controlled phase gate (and, briefly, a one-photon phase gate)
using the approximate Kerr nonlinearity provided by EIT. In this
paper, we rigorously analyze the dependence of the performance of
these gates on atomic dephasing and field detuning and intensity,
and we calculate the optimum parameters needed to apply a $\pi$
phase shift in a gate of a given fidelity. Although high-fidelity
gate operation will be difficult to achieve with realistic system
dephasing rates, the moderate fidelities that we believe will be
needed for few-qubit QIT seem much more obtainable.
\end{abstract}


\maketitle

\section{Introduction}

Quantum Information Science (QIS)\cite{Nielsen 2000} is a rapidly
emerging discipline with the potential to revolutionize
measurement, computation and communication. Sitting at the
intersection of quantum electronics, quantum optics, and
information theory, QIS offers a new paradigm for the collection,
transmission, reception, storage and processing of information,
based on the laws of quantum, rather than classical, physics.
Applications of QIS certainly have the potential to generate
totally new Quantum Information Technology (QIT), but ultimately
any future QIT industry will be justified by commercial interest
in the products and services that it supports. Practical results
in QIS are currently right at the cutting edge of experimental
quantum research, so the route to QIT is very hard going. From the
commercial perspective, a major challenge is the creation of
relatively simple QIT which is nevertheless economically viable.
This would generate revenue and expand current industrial
participation and interest in the field, effectively seeding a new
QIT industry, in the same way hearing aids were the first
commercial application of the transistor and the beginning of the
classical IT industry.

A long term goal for QIT is the realization of many-qubit scalable
quantum processors. It is known that these machines would
outperform their classical counterparts at certain tasks such as
factoring\cite{shor} and searching,\cite{grover} and the search
continues for new applications. A shorter term goal is the
realization of (say) 50-100 qubit processors. These would
certainly be better at quantum simulation\cite{lloyd} than any
foreseeable conventional IT and, as a research tool, would expose
QIT to a whole new class of curious and creative people, with all
their potential for new ideas and applications. However, perhaps
the most immediate QIT, that which will stimulate a new industry,
is based on or related to quantum communication\cite{gisin} and
metrology.\cite{lee}

It has become clear over the last two decades that computer and
network security that relies primarily on software protocols is
potentially porous, being based on unproven mathematical
assumptions. In QIS, quantum computation and communication
protocols can be devised in which unconditional privacy is
guaranteed by fundamental laws of physics. Although it is not
clear yet that quantum key distribution (QKD) will be the first
profitable application of QIT, it is possible that extensions of
QKD (e.g., controlled entanglement swapping), photonic state
comparison (for quantum signature verification), and full quantum
communication at high data rates will become compelling to
financial, medical, and other institutions and their customers.
Furthermore, it is already clear that distributed quantum
algorithms can efficiently enable solutions to economics problems
(e.g., public goods economics\cite{chen}) that are difficult to
treat with conventional mechanisms, but it is not yet known
whether other economic procedures---such as auctions---have
superior quantum solutions. Similarly, quantum metrology and
imaging have interest for the nanoscale manufacturing and physical
security industries, as these techniques allow tiny phase shifts,
displacements, and forces to be accurately measured remotely even
when the target is enclosed within an inaccessible or hostile
environment.

Although there are still open questions, one promising route for
starting a QIT industry is to found it on communication and
metrology applications, based on the generation, transmission,
processing, and detection of a few photonic qubits. It is
certainly clear that photons (or other quantum states of light)
are {\em the} qubits of choice for communication,\cite{gisin} and
so for few-qubit processing there is a case for keeping everything
optical. It is also the case that potentially useful processing
tasks could be performed with moderate (10--20\%) gate error
rates, rather larger than the stricter error bounds demanded for
fault-tolerant many-qubit processing. This ``all optical''
scenario is the motivation for our work. Clearly, taking this
approach, quantum information processing primitives based on
nonlinear quantum optics (such as a universal set of optical gates
and single-photon detectors) must be developed and fabricated.
These primitives would potentially allow the construction of
few-qubit nanoscale quantum optical processors that could be
incorporated into existing PCs and communication networks. In this
paper we discuss the possibility of realizing these primitives
through use of electromagnetically induced transparency (EIT).

\begin{figure}%[!htb]
\includegraphics[scale=0.5]{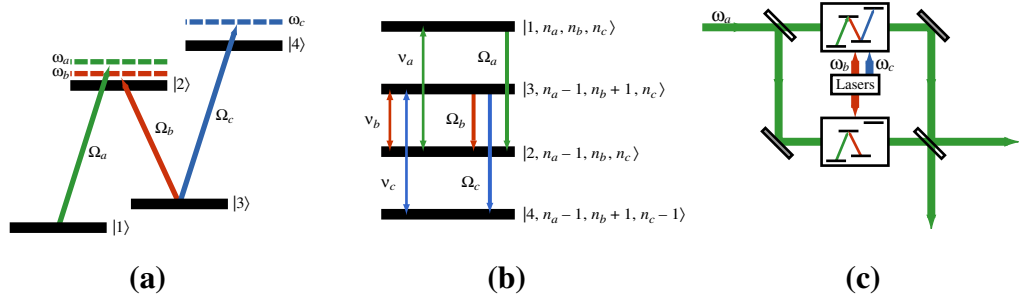}
\caption{ Electric dipole interaction
   between a four-level $\mathcal{N}$ atom and a nearly resonant
   three-frequency electromagnetic field. (a) In the semiclassical view, two atomic energy levels
   are separated by the energy $\hbar \omega_{ij}$,
   and coupled by a field oscillating at the frequency $\omega_k = \omega_{ij} +
   \nu_k$. The strength and phase of the corresponding dipole interaction is
   represented by the Rabi frequency $\Omega_k \propto
   \sqrt{n_k}$.
   (b) In the quantum view, the states of the atom + photons system
   separate into manifolds coupled internally by resonant
   transitions. (c) A model Mach-Zehnder interferometer
illustrating an architecture for a ``dual rail'' quantum
phase-shifter using four-level $\mathcal{N}$ atoms. The upper arm
is denoted by ``1'' and the lower arm by ``0.''} \label{fourlevel}
\end{figure}

 \section{Electromagnetically Induced Transparency}

In previous work,\cite{beau03} we considered a  model of the
nonlinear electric dipole interaction between three quantum
electromagnetic radiation fields with angular frequencies
$\omega_a$, $\omega_b$, $\omega_c$ and a corresponding four-level
$\mathcal{N}$ atomic system, as shown in Fig.~\ref{fourlevel}(a).
We considered $N$ atoms, fixed and stationary in a volume that is
small compared to the optical wavelengths, and we assumed that the
three frequency channels of the resonant four-level manifold of
the resulting quantum system shown in Fig.~\ref{fourlevel}(b) are
driven by Fock states containing $n_a$, $n_b$, and $n_c$ photons,
corresponding to the Rabi frequencies $\Omega_a$, $\Omega_b$, and
$\Omega_c$, respectively. As an example of the use of an EIT
system as a phase-shifter, we incorporate the atomic system into
the dual-rail Mach-Zehnder interferometer shown in
Fig.~\ref{fourlevel}(c). We wish to apply a phase shift to the
photon in mode $a$ on the upper rail, conditioned on the presence
of one or more photons in mode $c$. In one arm of the
interferometer, the four-level atoms are prepared using
$|\Omega_c| > 0$ to provide a phase shift at the probe frequency
$\omega_a$ while remaining largely transparent and dispersive. In
the second arm, $|\Omega_c| = 0$, and the system is tuned to match
the absorption and dispersion provided by the atoms in the first
arm, allowing the interferometer to remain time-synchronous.

We must be careful to demonstrate that the interaction of either
arm with a photon at the probe frequency that has entered the
interferometer at the input port will entangle the optical modes
with each other but \emph{not} with either collection of atoms.
Therefore, we solve the density matrix equation of motion in the
presence of a completely general Lindblad damping model, and
monitor the element $\rho_{10}(t)$ that corresponds to the initial
(ground) state of the entire collection of atoms in both
ensembles. If the field in mode $b$ is indeed described by a Fock
state, then using the quasi-steady-state approximation\footnote{In
Ref.~\cite{beau03}, we explicitly required $|\Omega_a| \ll
\gamma_{20}$ to obtain Eqs.~\ref{eq:rho10} and \ref{eq:W_10_4}.
However, our subsequent numerical work has shown us that those
results hold over a broad range of experimental parameters even
when $|\Omega_a| \approx \gamma_{20}$.} ($|\Omega_a| \lesssim
\gamma_{20}$) we obtain\cite{beau03}
 \begin{eqnarray} \label{eq:rho10}
\rho_{10}(t) \cong \rho_{10}(0)e^{(-\gamma_{10} + i W_{10})Nt},
 \end{eqnarray}
 where the Rabi frequencies $\Omega_k$ and detunings $\nu_k$ are
defined in Fig.~\ref{fourlevel}(a) and (b), and
 \begin{equation} \label{eq:W_10_4}
 W_{10} \equiv -\frac{\left[(\nu_a - \nu_b + i \gamma_{30})(\nu_a - \nu_b +
 \nu_c + i \gamma_{40}) - \left|\Omega_c\right|^2\right]\, \left|\Omega_a\right|^2}
 {(\nu_a + i \gamma_{20})\left[(\nu_a - \nu_b + i \gamma_{30})(\nu_a - \nu_b +
 \nu_c + i \gamma_{40}) - \left|\Omega_c\right|^2\right] -
 (\nu_a - \nu_b + \nu_c + i \gamma_{40}) \left|\Omega_b\right|^2}
 .
 \end{equation}
The constant $\gamma_{j0}, j \in \{2,3,4\}$ represents the net
decoherence rate (depopulation + dephasing) of level $j$ of the
quantum manifold shown in Fig.~\ref{fourlevel}(b) for the atom +
field system on the first rail relative to the evolution of a
system (absent mode $c$) on the second rail, while $\gamma_{10}$
represents the decoherence rate of the collective ground state
divided by the number of atoms $N$. Since the atomic levels
$|1\rangle$ and $|3\rangle$ are metastable by assumption, the
decoherence rates $\gamma_{10}$ and $\gamma_{30}$ represent pure
dephasing mechanisms.

 \section{Two-Qubit Phase Gates}

Suppose that we consider the concrete case of a phase gate that
couples single photons in modes $a$ and $c$ and is driven by a
coherent state in mode $b$. We wish to optimize the experimentally
controllable parameters so that we introduce a phase shift between
the two arms of the system with minimum error. We proceed by
choosing \emph{a priori} an error $\delta$ which occurs over the
entire gate operation, and determining the parameters needed for
the gate to perform with this level of error. There will be three
sources of error: the dephasing described by $\gamma_{10}$ and
$\gamma_{30}$; the additional depopulation described by
$\gamma_{20}$ and $\gamma_{40}$; and the error arising from the
finite value of $\alpha_b$ for the coherent state in mode $b$,
which prevents the system from evolving to a perfect phase-shifted
state even in the absence of decoherence.

Suppose that a phase shift of $-\phi \equiv
\operatorname{Re}\left(W_{10}\right) N t$ is applied by the phase
gate after a time $t$. Then the density matrix element given by
Eq.~(\ref{eq:rho10}) can be rewritten as $\rho_{10}(t) =
\rho_{10}(0)\, e^{-i \phi}\, e^{-\tau_\mathit{eff}}$, where the
effective decoherence time $\tau_\mathit{eff}$ is defined as
\begin{equation} \label{eq:gt}
\tau_\mathit{eff} \equiv -\frac{\gamma_{10} +
\operatorname{Im}\left(W_{10}\right)}{\operatorname{Re}\left(W_{10}\right)}
\, \phi .
\end{equation}
For a given gate fidelity $F$ (a measure of the distance between
the ideal output state of the gate and the state that is actually
generated\cite{Nielsen 2000,beau03}), the net error $\delta \equiv
1 - F^2$ due to dephasing and depopulation is $|\rho_{10}(0)|^2 (1
- e^{-2\tau_\mathit{eff}}) \approx \tau_\mathit{eff}/2$. In the
absence of significant dephasing, it is clear that a large value
of the detuning $\nu_c$ reduces the value of the ratio
$\operatorname{Im}\left(W_{10}\right)/\operatorname{Re}\left(W_{10}\right)$
and therefore $\tau_\mathit{eff}$; in fact, the maximum practical
value of $\nu_c$ would be set by the duration of the pulses in the
three electromagnetic modes and the experimental convenience of
accurately measuring the detuning. However, when the pure
dephasing terms $\gamma_{10}$ and $\gamma_{30}$ are finite,
$\tau_\mathit{eff}$ reaches a minimum for a finite value of
$\nu_c$. We assume that the dephasing rates are small enough that
$\gamma_{10}, \gamma_{30} \ll \gamma_{20}, \gamma_{40},
|\Omega_b|, |\Omega_c|$. Therefore, substituting
Eq.~(\ref{eq:W_10_4}) into Eq.~(\ref{eq:gt}), we obtain
 \begin{equation} \label{eq:nuc}
\nu_c = \left[\frac{\tilde{\gamma}_{20}\left(\gamma_{10}
\tilde{\gamma}_{20} +
\left|\Omega_a\right|^2\right)}{\tilde{\gamma}_{10}}\right]^\frac{1}{2}\,
\frac{Q_c^2}{Q_b^2}  \qquad \mathrm{and} \qquad \tau_\mathit{eff}
= 2 \left[\tilde{\gamma}_{10}\tilde{\gamma}_{20}\left(\gamma_{10}
\tilde{\gamma}_{20} +
\left|\Omega_a\right|^2\right)\right]^{\frac{1}{2}}
\frac{Q_b^2}{\left|\Omega_b\right|^2}
\frac{Q_c^2}{\left|\Omega_c\right|^2}
\frac{\phi}{\left|\Omega_a\right|^2} ,
 \end{equation}
where
 \begin{equation} \label{eq:g20_t}
Q_b^2 \equiv \gamma_{20} \gamma_{30} + |\Omega_b|^2, \qquad Q_c^2
\equiv \gamma_{30} \gamma_{40} + |\Omega_c|^2, \qquad
\tilde{\gamma}_{10} \equiv \gamma_{10} +
\gamma_{30}\frac{\left|\Omega_a\right|^2 }{Q_b^2}, \qquad
\mathrm{and} \qquad \tilde{\gamma}_{20} \equiv \gamma_{20} +
\gamma_{40}\frac{\left|\Omega_b\right|^2 }{Q_c^2} .
 \end{equation}

Consider the case of a $\pi$ phase gate, and assume that the pure
dephasing rates are small enough that $\gamma_{20} \gamma_{30} \ll
|\Omega_b|^2$, $\gamma_{30} \gamma_{40} \ll |\Omega_c|^2$, and
$|\Omega_a|^2 \ll |\Omega_b|^2$. We see immediately that we can
have $\tau_\mathit{eff} \ll 1$ only if $\gamma_{10}
\tilde{\gamma}_{20} \ll |\Omega_a|^2$, giving $\nu_c \approx
\sqrt{\tilde{\gamma}_{20}/\tilde{\gamma_{10}}}
(|\Omega_c|^2/|\Omega_b|^2) |\Omega_a|$ and $\tau_\mathit{eff}
\approx 2 \pi \sqrt{\tilde{\gamma_{10}}
\tilde{\gamma}_{20}}/|\Omega_a|$. Now, if $\gamma_{20} \approx
\gamma_{40}$, then we must have $\gamma_{10} \ll |\Omega_a|^2
|\Omega_c|^2/\gamma_{20} |\Omega_b|^2$, which could be difficult
to achieve in a realistic experiment. However, if we can embed the
atomic system in a photonic crystal and suppress spontaneous
emission from atomic state $|4\rangle$,\cite{beau03} this
requirement is eased considerably to $\gamma_{10} \ll
|\Omega_a|^2/\gamma_{20}$.

In addition to the decoherence error, there is also an error
introduced when the Fock state $|n_b\rangle$ is replaced by the
coherent state $|\alpha_b\rangle$ with $\langle n_b \rangle =
|\alpha_b|^2$. When dephasing can be completely neglected,
Eq.~(\ref{eq:W_10_4}) shows that $W_{10} \propto n_a n_c/n_b$, and
it is clear that the gate performs a phase shift when there is a
photon present in \emph{both} modes $a$ and $c$, and no phase
shift when either is absent. However, mode $b$ is driven by a
coherent state, and an input coherent state will evolve according
to\cite{beau03}
 \begin{equation} \label{eq:psi_prime}
 \ket{\psi(t)} = e^{-\half \left|\alpha_b\right|^2} \sum_{n_b = 0}^{\infty}
\frac{\alpha_b^{n_b}}{\sqrt{n_b!}}e^{-i\, n_a n_c \phi(t)\,
\left|\alpha_b\right|^2 /n_b} \ket{\{1\}, n_a, n_b, n_c} ,
 \end{equation}
where $\phi(t) \equiv \tilde{W} t$, $\tilde{W} \equiv N
|\tilde{\Omega}_a|^2 |\tilde{\Omega}_c|^2/\nu_c
|\tilde{\Omega}_b|^2 |\alpha_b|^2$, and $|\tilde{\Omega}_k| \equiv
|\Omega_k|/\sqrt{n_k}$. Clearly, if $|\alpha_b| \gg 1$, then only
terms with $n_b \approx |\alpha_b|^2$ make a significant
contribution to the sum, and the phase can be approximately
factored out as $\ket{\psi(t)} \approx e^{i \phi(t)}
\ket{1,n_a,\alpha_b,n_c}$. However, if the coherent state is too
weak, then the phase shift will be imperfectly implemented with a
significant error, even in the absence of decoherence.
Furthermore, we see from Eqs.~(\ref{eq:gt}) and (\ref{eq:g20_t})
that the effect of the decoherence increases with the intensity of
the coherent state, and therefore for a given dephasing there is
some ideal value of $\alpha_b$ that minimizes the
error.\cite{munr03}

\begin{figure}
\includegraphics[scale=0.6]{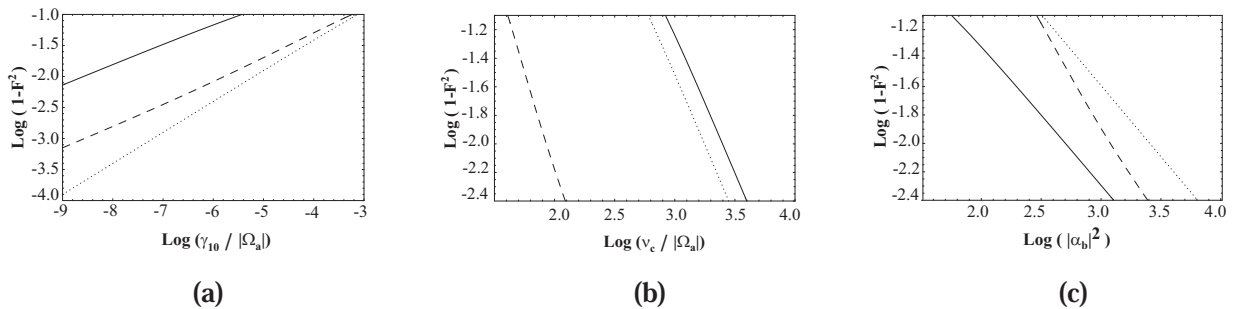}
\caption{The gate error as a function of (a) the logarithm of the
dephasing, (b) detuning, and (c) coherent state intensity required
to generate that error. In all plots, the solid lines (---) shows
the error when emission from level $|4\rangle$ is not suppressed,
i.e. $\gamma_{40} \sim \gamma_{20}$. The dashed lines (- - -)
illustrates the case when $\gamma_{40}$ has been suppressed
relative to $\gamma_{20}$ by a factor of 1000. The dotted lines
($\cdots$) give the error in the case of the one-qubit phase gate
when modes $b$ and $c$ are occupied by coherent states. In (c),
the solid and dashed lines correspond to an ideal value of
$|\alpha_b|$. However, the dotted line ($\cdots$), shows a
\emph{minimum} value that $|\alpha_b|$ must take for the error in
the one-qubit gate to remain below the value shown; $|\alpha_b|$
can take on larger values without affecting either the decoherence
or the value of $\nu_c$ required to minimize the error. For the
one qubit gate, $|\alpha_c| \gg |\alpha_b|$.} \label{fig:gplot}
\end{figure}

We are now ready to estimate the magnitude of the dephasing that
must be achieved to obtain a given error. We assume that
$|\Omega_a| \approx \gamma_{20}$ and $|\tilde{\Omega}_b|^2 \approx
|\tilde{\Omega}_c|^2$, and we consider the most pessimistic case
where $\gamma_{20} \approx \gamma_{40}$. We obtain
\begin{equation} \label{eq:dp}
\frac{\gamma_{10}}{|\tilde{\Omega}_a|} = \left( \frac{\delta}{\phi
} \right)^2\frac{1}{n_b} .
\end{equation}
This equation gives an estimate of the error arising from a
\emph{Fock} input state with an optimized value of $\nu_c$ for a
particular level of decoherence. In Fig.~\ref{fig:gplot}(a) we
plot the error induced on a \emph{coherent} input state due to
both decoherence and the finite intensity of the coherent state as
a function of the dephasing $\gamma_{10}$, where we have chosen
both $\nu_c$ and $|\alpha_b|$ to minimize the error. Figures
\ref{fig:gplot}(b) and \ref{fig:gplot}(c) shows these values of
$\alpha_b$ and $\nu_c$ chosen to give the largest possible
dephasing for a given error.

As noted above, if $\gamma_{40}$ can be suppressed, then the
appropriate constraint $\gamma_{10} \ll |\Omega_a|^2/\gamma_{20}
\approx \gamma_{20}$ becomes relatively easy to satisfy. For the
requirement $|\alpha_b|^2 \ll \gamma_{40}/\gamma_{20}$, the
dephasing can be larger than in the unsuppressed case by a factor
of $|\alpha_b|^2$ and still produce the same error. By contrast,
for $|\alpha_b|^2 \gg \gamma_{40}/\gamma_{20}$, the dephasing can
be larger by a factor of $\gamma_{40}/\gamma_{20}$. Therefore,
suppressing $\gamma_{40}$ means that the detuning required is also
much smaller.

In order to perform a two-qubit conditional phase gate with an
error of 20\% (and assuming we cannot suppress the depopulation
from level $|4\rangle$), from Fig.~\ref{fig:gplot} we find that
the required parameters are $|\alpha_b| = 10$,
$\nu_c/|\tilde{\Omega}_a| = 125$, and
$\gamma_{10}/|\tilde{\Omega}_a| = 6 \times 10^{-7}$, and the gate
operation time is determined by $|\Omega_a| N t = 1.25 \times 10^4
\phi$. However, if $\gamma_{40}$ can be suppressed by a factor of
1000 relative to $\gamma_{20}$, the requirements are much reduced:
the dephasing needed is $\gamma_{10}/|\tilde{\Omega}_a| = 2 \times
10^{-4}$, the detuning is $\nu_c/|\tilde{\Omega}_a| = 30$, and the
coherent state must have $|\alpha_b| = 20$. The corresponding gate
operation time is determined by $|\Omega_a| N t = 160 \phi$.

This system can also be used as a one-qubit phase shift gate (or
an EIT-based QND detector,\cite{munr03} where we can assume that
$|\Omega_c| \approx |\Omega_b|$), \emph{without} the extra effort
described above for suppression of spontaneous emission from the
atomic level $\ket{4}$. If mode $c$ remains in the single-photon
Fock state, the system will act as a phase shifter on mode $a$,
and the above analysis applies. Alternatively, we can replace the
quantum state in mode $c$ with an intense coherent-state driving
field. Now moderate (i.e., non-classical) values of $|\alpha_b|$
and $|\alpha_c|$ introduce errors. However, examining
Eqs.~(\ref{eq:g20_t}), we see that---unlike for the two-qubit
gate---the effects of decoherence depend much less sensitively on
the intensities of the coherent states. As shown in
Fig.~\ref{fig:gplot}, this means that we can eliminate the error
due to the finite size of the coherent state, and we are only left
with an error due to decoherence.

\section{Conclusion}

We have studied gates for quantum information processing based on
a quantum treatment of EIT systems. We have analyzed in detail the
performance of a two-qubit phase gate (and, by extension, that of
a one-qubit phase gate) as functions of both atom and field
properties, and we have described a general optimization method
that selects a detuning that minimizes the gate error for a given
phase shift. The resulting constraints on the allowable dephasing
rates in these systems are quite stringent for high-fidelity
operation. However, if the spontaneous emission rate from the
atomic level $|4\rangle|$ can be suppressed significantly, then
demonstration of a moderate-fidelity phase gate becomes
experimentally achievable.

\end{document}